\documentclass[12pt]{article}
\usepackage{amssymb,amsmath,epsfig}
\begin{document} \parindent=0pt
\parskip=6pt \rm

 \begin{center}
 {\bf \Large About phase transitions in Bose gases
 at constant density and constant pressure}

 \vspace{0.5cm}

{\bf Velin G. Ivanov, Dimo I. Uzunov}

 {\em CP Laboratory, G. Nadjakov
Institute of Solid State Physics,\\
 Bulgarian Academy of Sciences, BG-1984 Sofia, Bulgaria.}

 \end{center}

\begin{abstract}
The phase transitions in Bose gases at constant volume and constant
pressure are considered. New results for the chemical potential, the
effective Landau-Ginzburg free energy and the equation of state of the
Bose-Einstein condensate in ideal Bose gases with a general form of the
energy spectrum are presented. Unresolved problems are discussed.
\end{abstract}

{\bf 1. Introduction}

The Bose-Einstein condensation (BEC) is a phenomenon due to the quantum
statistical correlations (pseudo-interactions) in ideal Bose gases
(IBG) of non-interacting bosons. This is a condensation in the momentum
space ($\hbar\vec{k}$) but also it possesses some features of the usual
(Van der Waals) condensation~\cite{Huang:1963, Landau:1980,
Uzunov:1993, Shopova:2003}. This phenomenon strongly depends on the
spatial dimensionality $d$, and the energy spectrum $\varepsilon(k) =
(\hbar^2k^{\sigma}/2m)$ of the bosons [$0< \sigma \leq 2$; $k
=|\vec{k}|;$ $\vec{k} =\{ k_j= (2\pi/n_j/L_j)\}$, where $j=1,...,d;$ $
n_j = 0,\pm 1,...$] in IBG in volume $V=(L_1...L_d) \sim L^d$ and
periodic boundary conditions ~\cite{Uzunov:1993, Shopova:2003}.

The original Bose-Einstein condensate (BEC) is not superfluid but the
latter state is possible in many-body systems of interacting bosons
(nonideal Bose gas, or, shortly, NBG)~\cite{Huang:1963,Landau:1980,
Uzunov:1993, Shopova:2003, Tilley:1974}. The phenomena of BEC and
superfluidity have a number of similar features and both of them are
widely discussed in various problems of astrophysics (see, e.g.,
Refs.~\cite{Tilley:1974, Celebonovic:2000, Tsuruta:1998, Link:2003}. In
particular, BEC and superfluidity are relevant to the treatment of the
so-called plasma-solid transition in ``astrophysical
matter''~\cite{Celebonovic:2000}, and for the behaviour of the neutron
component of the dense matter in the interior of neutron
stars~\cite{Tilley:1974, Tsuruta:1998, Link:2003, Rojas:2004}; for
applications to cosmological models of dark energy and dark matter, see
Ref.~\cite{Nishiyama:2004}.

The main properties of BEC are known (see, e.g.,
Ref.~\cite{Shopova:2003, Gunton:1968, Cooper:1968, Gunther:1974,
Lacour:1974, Busiello:1985}. Recently, BEC at constant density and
constant pressure has been reviewed in Ref.~\cite{Ivanov:2004}. Here we
shall discuss the equation of state of IBG and BEC for spinless bosons.
The treatment can be generalized for bosons with
spin~\cite{Yamada:1982, Frota:1984}.

The superfluidity and the effect of interparticle interactions on BEC
can be treated by both the mean-field like
Gross-Pitaevskii~\cite{Dalfovo:1999} and the microscopic
Beliaev-Popov~\cite{Popov:1983} approaches. We shall discuss the latter
within the renormalization group theory~\cite{Uzunov:1993} and for this
reason we shall consider the following action of NBG~\cite{Uzunov:1993,
Shopova:2003, Popov:1983}:
\begin{equation}
{\cal{S}}[\phi]  =  -\sum_{q}G^{-1}_0(q)|\psi(q)|^2 - \frac{u}{2\beta
N}\sum_{q_1,q_2,q_3} \psi^{\ast}(q_1)\psi^{\ast}(q_2)\psi(q_3)\psi(q_1
+ q_2 - q_3)\;,
\end{equation}
where $q=(\omega_l, \vec{k})$ is a $(d+1)$-dimensional
frequency-momentum vector, $\omega_l = 2\pi l k_BT/\hbar$ is the
(Bose-)Matsubara frequency ($l = 0,\pm 1,...)$, $\psi(q)$ is a C-number
Bose field, $u $ is the interaction constant, $N$ is the number of
particle and $\beta = 1/k_BT$. The (bare) correlation (Green) function
$G_0(q) = \langle |\psi(q)|^2\rangle_0$ is given by
\begin{equation}
 G_0^{-1}(q) \;  = \; i\omega_l + \varepsilon(k) + r\:,
\end{equation}
where $(-r) = \mu \leq 0$ is the chemical potential of IBG ($u=0$).

Note, that the thermodynamics of NBG is given by the grand canonical
thermodynamic potential $\Omega (T,V,\mu) =
-\beta^{-1}{\mbox{{\cal{Z}}}}$, where the grand canonical partition
function ${\cal{Z}}(T,V,\mu)$ is defined by ${\cal{Z}} =
\int{\cal{D}}\phi\:\mbox{exp}\:({\cal{S}})$ - a functional integral
over the possible field configurations. The usual field theoretical
investigations of the action (1) are performed in the thermodynamic
limit: $N\rightarrow \infty$, $V \rightarrow \infty$, provided $0 \leq
\rho = (N/V) < \infty$.

Using the models of IBG ($v \equiv 0$) and NBG defined by
Eqs.~(1)~--~(2) we shall consider the equation of state for BEC. For
IBG this can be performed exactly, whereas for NBG we must use the loop
expansion (see, e.g., Ref.~\cite{Uzunov:1993}). The main features of
BEC can be revealed within the one-loop
approximation~\cite{Uzunov:1993, Toyoda:1982}. We shall discuss the
effect of the inter-particle interaction on BEC by using results from
preceding works.

 We wish to emphasize that the phase transition to BEC cannot be
easily put to the usual classification of phase
transitions~\cite{Uzunov:1993}. Above $T_c$ this phase transition
resembles certain features of second order phase transitions but is
rather different from the standard notion about these
transitions~\cite{Uzunov:1993}. On the other side, the equation of
state of IBG below $T_c$ is quite similar to known equations of
(almost) first-order phase transitions and tricritical points
~\cite{Uzunov:1993}. There is a close similarity between the phase
transition properties of IBG and the spherical model in the
ferromagnetism, and this point will be discussed in the remainder of
this report. In our consideration we shall essentially use results from
preceding works~\cite{Ivanov:2004,Gunton:1968, Cooper:1968,
Gunther:1974, Lacour:1974, Busiello:1985} (for a comprehensive
reference to original papers, see, e.g., the reviews~\cite{Uzunov:1993,
Shopova:2003}). We work with general values of $d$ and $\sigma$ but
 our consideration includes the important case of $d=3$,
$\sigma =2$.

{\bf 2. Free energy and thermodynamics}

For a convenience we shall introduce a fictitious external field $h$
which is thermodynamically conjugated to the order parameter $\Psi=
\langle \psi(0) \rangle$ of uniform BEC. Note, that in case of uniform
BEC the $(q=0$)-mode $\psi(0)$ can be represented by the sum $\psi(0) =
\langle \psi(0)\rangle + \delta\psi(0)$, where $\delta\psi(0)$ is the
(uniform) fluctuation mode. The consideration of an external uniform
field $h$ can be performed by adding a term
\begin{equation}
{\cal{S}}_h = \frac{1}{N}\left[h\psi^{\ast}(0) + c.c.)\right]\:.
\end{equation}

The thermodynamic potential of IBG can be written in the form
\begin{equation}
\Omega
(T,r,h)=-\beta^{-1}V\lambda_T^{-d}A(d,\sigma)g_{d/{\sigma}+1}(\beta r
)- \frac{1}{N}\frac{hh^{*}}{r},\hspace{0.5cm}
\end{equation}
where $g_{\nu}(y)$ is the Bose function~\cite{Uzunov:1993,
Busiello:1985, Frota:1984}, the thermal wavelength is given by
\begin{equation}
\lambda_T=\left( {2\pi \hbar^2\over{mk_BT}} \right)^{1/{\sigma}}\:,
\end{equation}
 and
\begin{equation}
A(d,\sigma)={2^{1-d+2d/{\sigma}}\Gamma(d/{\sigma})\over{\sigma\pi^{d(1/2-1/{\sigma})}
\Gamma(d/2)}},\quad A(d,2)=1.
\end{equation}
The potential (4) obeys the differential relation $d\Omega=-SdT+ N
d{r}-\Psi dh^{*}-\Psi^{*} dh$.

By a suitable Legendre transformation we obtain another thermodynamic
potential $\tilde{\Omega}(T,r,\Psi)$, where the natural variable is
$\Psi$:
\begin{equation}
\widetilde{\Omega}=-\beta^{-1}V\lambda_T^{-d}A(d,\sigma)g_{d/{\sigma}+1}(\beta
r )+Nr{\Psi}^2.
\end{equation}
We can restrict ourselves, without a loss of generality, to real $h$
and $\Psi$. The susceptibility is then $\chi_{T}=\partial\Psi/\partial
h=1/Nr\sim t^{-\gamma}$, where $\gamma$ is the susceptibility critical
exponent and $t=(T-T_c)/T_c$ (see,e.g., Ref.~\cite{Uzunov:1993}).

For small $r$ and energies ($\varepsilon \ll k_BT$), which correspond
to the critical regime near the phase transition point, the correlation
function $\chi(k) = G_0(0,\vec{k})$ takes the form $\chi(k)^{-1}  \sim
(ck^{\sigma} + r)$. This gives us the correlation length
$\xi=(c/r)^{1/\sigma}\sim t^{-\nu}$ and $\chi(k)\sim k^{-2+\eta}$ at
$r=0$. From $\chi(k) \sim k^{\sigma}$ we obtain that the Fisher
exponent $\eta$, defined by $\chi(k) \sim k^{-2 + \eta}$, is equal to
$(2-\sigma)$ for all dimensional ranges and possible constraints (of
constant volume $V$ or constant pressure $P$).

In order to obtain the correlation length exponent $\nu$ and the
exponent $\gamma$ of the susceptibility $\chi_T$, we need to calculate
the function $r(t)$. The latter is different for the cases $V= const$
and $P=const$. In the thermodynamic limit, the constraint of constant
volume is equivalent to a constraint of constant density ($\rho =
const$). The results for the critical exponents corresponding to these
cases are summarized in Ref.~\cite{Ivanov:2004} for various spatial
dimensions $d$. Here we restrict our consideration of the critical
exponents and the equation of state to features which demonstrate the
difference between the phase transition to BEC and both first and
second order phase transitions for the most interesting interval of
spatial dimensions $\sigma < d < 2\sigma$ that includes the case
$\sigma =2$, $d=3$.\\

{\bf 3. Constant density}

To consider the effect of the thermodynamic condition of constant
density $\rho = (N/V)$ we need the free energy $F =Vf(T,\rho,\Psi)$. To
obtain this thermodynamic potential near the phase transition point to
BEC we expand the Bose function in (4) in powers of $\beta r$ and
express $r$ as a function of $\rho$. This procedure is performed for
the potential $\tilde{\Omega}$ given by Eq.~(7). Further we obtain the
potential $F$ with the help of a Legendre transformation of the form:
\begin{equation}
F=\tilde{\Omega}-Nr|_{r=r(\rho)}\:,
\end{equation}
\begin{equation}
\rho=\frac{\partial \widetilde{\Omega}}{V\partial
r}=\lambda_T^{-d}A(d,\sigma)g_{d/{\sigma}}(\beta r )+\rho {\Psi}^2.
\end{equation}

The lowest temperature $T_c$, at which the system does not condensate
($\Psi =0$) is obtained from Eq.~(9). When $d\leq \sigma$, the Bose
function is divergent for $(\beta r) \rightarrow 0$ which means that
BEC may occur only at the absolute zero $(T_c = 0)$ - a zero
temparature BEC~\cite{Busiello:1985}. When $d > \sigma$, we obtain
that~\cite{Shopova:2003, Lacour:1974}
\begin{equation}
T_c(\rho) =
\frac{2\pi\hbar^2}{mk_B}\left[\frac{\rho}{A(d,\sigma)\zeta(d/\sigma)}\right]^{\sigma/d}
> 0 ,
\end{equation}
where $g_{d/\sigma}(0) = \zeta(d/\sigma)$ is the zeta function. Below
the critical temperature $T_c$, BEC occurs $(\Psi > 0)$.

 For the most interesting case of dimensions $\sigma < d <2\sigma$,
the result for the free energy density $ f(T,\rho, \Psi)$ to the lowest
order in $|t| \ll 1$ is given by:
\begin{equation}
 f = C_f\left(\Psi^2 + \frac{d}{\sigma}\:t\right)^{d/(d-\sigma)},
\end{equation}
where
\begin{equation}
C_f =
\left(\frac{d}{\sigma}-1\right)\left[\frac{\zeta(d/\sigma)}{|\Gamma(1-d/\sigma)|}
\right]^{\sigma/(d-\sigma)}(k_BT_c)\rho\:.
\end{equation}
In our derivation of the energy (11) a $\Psi$-independent term has been
neglected. Such terms can be ignored because they belong to the energy
of the disordered phase ($\Psi = 0$). For this reason the net free
energy of the BEC can be obtained from (11) by neglecting the
$\Psi$-independent term of type $t^{d/(d-\sigma)}$. The free energy
(11) is of Landau-Ginzburg type~\cite{Uzunov:1993}; one may easily
expand $f(\Psi)$ in powers of $\Psi$.

The external field $h$ is fictitious and can be neglected in studies of
the thermodynamics. In this case, the equation of state $h \sim
\partial (f/\partial \Psi)$ becomes $(\partial f/\partial \Psi) = 0$.
From Eq.~(11) we easily obtain the equation of state for $h=0$:
\begin{equation}
\Psi \left(\Psi^2 + \frac{d}{\sigma}\:t\right)^{\sigma/(d-\sigma)} = 0.
\end{equation}

The solutions of Eq.~(13) are: $\Psi = 0$ (disordered phase), and
$\Psi^2 = -(d/\sigma)t > 0$ (corresponding to BEC for $t<0$). Note,
that the chemical potential ($\mu = -r$) can be obtained in the form
\begin{equation}
\mu = -
k_BT\left[\frac{\zeta(d/\sigma)}{|\Gamma(1-d/\sigma)|}\right]^{\sigma/(d-\sigma)}
(\Psi^2 + \frac{d}{\sigma}\:t)^{\sigma/(d-\sigma)}.
\end{equation}
BEC is possible for $t<0$ under the condition $\mu = 0 $. The latter
introduces thermodynamically forbidden (unstable) domains in the phase
diagram.

The main conclusion which can be drawn from the Ginzburg-Landau free
energy (11) of IBG at constant $\rho$ is that the point $t = 0$
resembles a trictitical point~\cite{Uzunov:1993}. Usually such
multicritical points occur on a phase transition line where the phase
transition changes from second order to a symmetry conserving first
order phase transition (or vice versa). Here this is not the case but
the similarity with the usual tricriticality is in the fact that the
coefficients of both the $\Psi^2-$ and the $\Psi^4-$terms in (11) tend
to zero for $t \rightarrow 0$. On the other side the phase transition
to BEC is a continuous phase transition which exhibits critical
exponents identical to the critical exponents known from the Berlin-Kac
spherical model in ferromagnetism~\cite{Berlin:1952,Langer:1963} (see
also Refs.~\cite{Uzunov:1993,Shopova:2003}). Here the condition of
constant density $\rho$ plays the role of the spherical condition for
the spins in the spherical model.

{\em Interaction effect}. Most of these exceptional properties of the
phase transition to BEC in IBG are not present in systems of
interacting bosons described by NBG. In this case an additional term of
type $u\Psi^4$ will appear in the effective free energy
$f(T,\rho,\Psi)$ and the respective coefficient of the $\Psi^4-$ term
will remain finite at $T_{\lambda}$ -- the critical temperature to
superfluid state in interacting Bose fluids (gases and liquids). The
generalization of our treatment to the case of interacting bosons
described by the action (1) may lead to new thermodynamic properties. A
similar generalization for the spherical model was made in
Ref.~\cite{Langer:1963}. Another way of treatment of NBG may be
performed within the loop expansion~\cite{Toyoda:1982}. It is believed
that the $\lambda-$transition described by NBG belongs to the so-called
XY universality class of standard second order phase
transitions~\cite{Uzunov:1993}.

Recent studies of interacting bosons indicated discrepancies in the
theoretically predicted values of the critical temperature
$T_{\lambda}$~\cite{Baym:1999, Huang:1999, Schakel:2000, Schakel:2003}.
The problem for the calculation of the phase transition temperature of
interacting many-body systems is a hard and still unresolved problem of
the theory~\cite{Uzunov:1993}. The fluctuation shifts of the critical
temperature that are usually calculated from field models (see, e.g.,
Ref.~\cite{Uzunov:1993} are very small and do not include essential
contributions due to the large-momentum (high-energy) fluctuations of
the order parameter field $\psi$. Therefore, the correct treatment of
the phase transition temperature in many-body systems with
interparticle interactions requires new theoretical methods.

{\bf 4. Constant pressure}

The only papers where the effect of the constraint of constant pressure
on the phase transition properties of IBG has been investigated so far
are Refs.~\cite{Gunther:1974, Lacour:1974, Shopova:2003, Ivanov:2004}.
This case should be taken in mind in interpretations of real
experiments, in particular, in low-temperature experiments on BEC in
trapped atomic gases (see. e.g., Ref.~\cite{Dalfovo:1999}), where the
density $\rho$ varies but the pressure $P$ is (almost) fixed. In this
case BEC occurs at finite temperatures ($T_c > T > 0$) for all spatial
dimensions $d>0$. The critical temperature will be~\cite{Shopova:2003}
\begin{equation}
T_c(P) = \left[\frac{\lambda_0^d P}{\zeta(1 +
d/\sigma)A(d,\sigma)k_B}\right]^{\sigma/(d + \sigma)},
\end{equation}
where $\lambda_0$ is given by $\lambda_0 = \lambda T^{1/\sigma}$ and
Eq.~(5). For $d=\sigma=2$ one obtains the result for $T_c$ known from
Ref.~\cite{Gunther:1974}.

Although a number of results are known from preceding works, in
particular, for the two-dimensional case ($d=2$)~\cite{Gunther:1974},
the entire picture of the phase transition to BEC at constant pressure
is not still clear for all spatial dimensions $d$. In particular, this
is the case of interacting bosons. We have no information about
research papers devoted to this problem.

{\bf 5. Concluding remarks}

We have presented a brief discussion of several properties of the phase
transition to BEC. New results have been obtained for the effective
Landau-Ginzburg free energy (11), the equation of state (13) and the
chemical potential (14) of BEC in ING for general values of $d$ and
$\sigma$.

The effects of the constraints of constant density and constant
pressure on BEC in ING and the condensation to a superfluid phase in
systems of interacting bosons (NBG) are not yet clarified in a
comprehensive way. The renormalization group methods~\cite{Uzunov:1993,
Shopova:2003} reveal a dimensional (quantum to classical crossover) in
interacting Bose systems and other basic universality features of the
so-called quantum phase transitions at zero and very low temperatures
but the equation of state below the phase transition point is not
investigated in details. This is the situation for the whole variety of
Bose systems known in condensed matter physics~\cite{Uzunov:1993,
Shopova:2003} and, in particular, for Bose fluids of real atoms.

For Bose fluids the concept of universality of the quantum critical
phenomena seems to be invalid and mainly for this reason the quantum
phase transitions of second order can be classified in two groups: as
universal and non-universal (see Ref.~\cite{Shopova:2003}). However, a
number of examples indicate that the quantum phase transitions are
often of first order or are described by multicritical points which are
different from the critical points of standard second order phase
transitions.

In astrophysics, a quantum phase transition ($T_c \sim 0$) may occur in
cases of very low density or very low pressure of the respective Bose
fluid; see Eqs.~(10) and (14). For the matter in the interior of
neutron stars one should investigate the so-called classical
limit~\cite{Shopova:2003} in which the quantum fluctuations are
irrelevant. This is the case of $T_c>0$ discussed in the prevailing
part of our report.

BEC of spin bosons~\cite{Yamada:1982} can be treated within the
framework of a generalization of our treatment. For this aim one may
consider a complex vector field $\vec{\psi}(q) = \left\{\psi_{\alpha};
\alpha = 1,...n/2\right\}$. For $n=2$ one obtains the complex scalar
field in Eq.~(1). When such a system is placed in an external magnetic
field, essentially new results can be obtained~\cite{Yamada:1982}. The
unresolved problem about the superfluidity of interacting spin bosons
in external magnetic field is of special interest in astrophysics.

\end{document}